\documentclass{mem}
\usepackage{natbib}
\usepackage{txfonts}
\usepackage{balance}
\usepackage{graphicx}
\usepackage{flushend}
\usepackage[breaklinks,pdftex]{hyperref}
\idline{75}{282}
\begin{document}
\def\teff{$T\rm_{eff }$}
\def\kms{$\mathrm {km s}^{-1}$}
\def\ms{M$_\odot$}
\def\rr{RR$_{0.61}$}
\def\rrl{RR$_{0.68}$}

\title{
Blazhko effect and the Petersen diagram
}

   \subtitle{}

\author{
H. \,Netzel\inst{1,2,3} 
          }

\institute{
Konkoly Observatory, Research Centre for Astronomy and Earth Sciences, E\"otv\"os Lor\'and Research Network; MTA Centre of Excellence (ELKH),\\ H-1121 Konkoly Thege Mikl\'os \'ut 15-17, Budapest, Hungary
\and
MTA CSFK Lend\"ulet Near-Field Cosmology Research Group\\ H-1121 Konkoly Thege Mikl\'os \'ut 15-17, Budapest, Hungary
\and
ELTE E\"otv\"os Lor\'and University, Gothard Astrophysical Observatory, Szent Imre h. u. 112, 9700, Szombathely, Hungary\\
\email{henia@netzel.pl}
}

\authorrunning{Netzel}

\titlerunning{Variable stars}

\date{Received: Day Month Year; Accepted: Day Month Year}

\abstract{
Pulsations in RR Lyrae stars and classical Cepheids were thought to be relatively
simple since they typically pulsate only in one or two radial modes. This picture
changes at a closer look when modulation or additional low-amplitude signals are
detected. I will review different multi-periodic groups known among classical
pulsators, including stars showing the Blazhko modulation.
}
\maketitle{}

\section{Introduction}

\begin{figure*}
    \centering
    \includegraphics[width=0.9\textwidth]{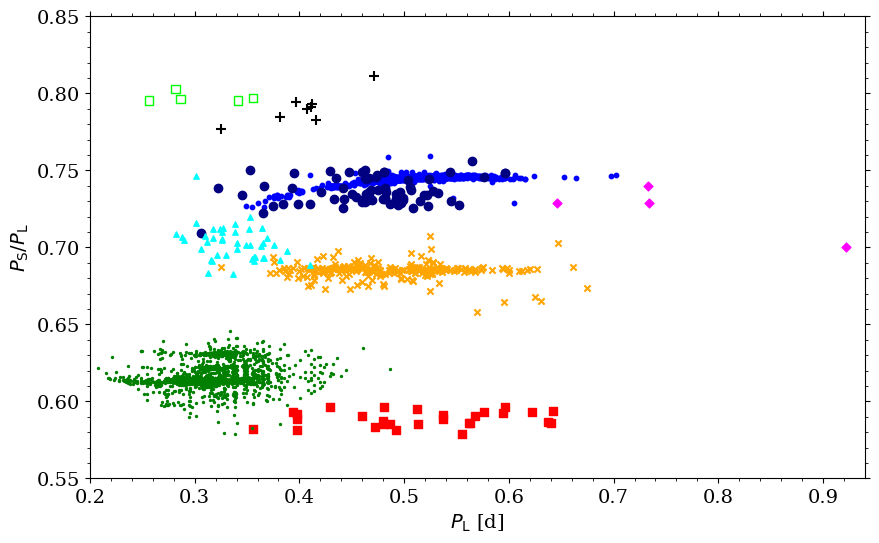}
    \caption{Petersen diagram for multi-mode RR Lyrae stars.}
    \label{fig:pet_rrl_all}
\end{figure*}

For many years RR Lyrae stars and classical Cepheids were considered very simple pulsators. The majority of them pulsate in one or two radial modes. The only mystery was related to some RR Lyrae stars that show quasi-periodic modulation of pulsations known as the Blazhko effect \cite{blazhko}. However, the past decade revolutionized our view on these relatively simple pulsators. Thanks to large-scale ground-based surveys and space-based missions, many low-amplitude signals were detected in RR Lyrae stars and classical Cepheids. New distinct multi-periodic groups emerged in the Petersen diagram, which is a diagram of period ratio vs longer period or its logarithm (see Figures~\ref{fig:pet_rrl_all} and \ref{fig:pet_cep_all}).

Here we review the currently known groups of multi-mode classical pulsators. We also discuss the advances in the long-standing problem of the Blazhko effect. 

\section{Petersen diagram for RR Lyrae stars}\label{sec:rrl}

Petersen diagram for RR Lyrae stars is presented in Fig.~\ref{fig:pet_rrl_all}.

\subsection{Multi-mode radial pulsations}
The simplest multi-periodic pulsations are double-mode pulsations in radial modes. Double-mode pulsations in fundamental mode and first overtone (RRd) were detected for the first time in AQ Leo by \cite{jerzykiewicz1977}. Since then, many more such stars were discovered in various stellar systems. RRd stars are plotted in the Petersen diagram in Fig.~\ref{fig:pet_rrl_all} with blue circles. They form a characteristic progression from 0.72 (short periods) to 0.75 (long periods).

Related to the RRd stars described above are the anomalous RRd stars (aRRd), which are plotted with navy circles in Fig.~\ref{fig:pet_rrl_all}. They were detected in multiple systems, e.g. in LMC and SMC \citep{soszynski2016}, in the globular cluster M3 \citep{jurcsik2014}, and in the Galactic field \citep{smolec2015}. They differ from classical RRd stars in a few ways. First, they show atypical period ratios and are located above or below the classical RRd progression in the Petersen diagram. In contrast to RRd stars, in aRRd stars, the fundamental mode has a higher amplitude than the first overtone. Moreover, many aRRd stars show the Blazhko modulation, which will be discussed in Sec.~\ref{sec:blazhko}.

Less common are double-mode RR Lyrae stars pulsating in the fundamental mode and second overtone. They are plotted with red squares in Fig.~\ref{fig:pet_rrl_all}. The second overtone is detected typically based on the space-based data \citep{jurcsik2008,poretti2010,chadid2010,benko2010,benko2014}.

Very rare objects are triple-mode RR Lyrae stars. All of them show pulsations in the fundamental, first and second overtones. The period ratio formed by first and second overtones is plotted in Fig.~\ref{fig:pet_rrl_all} with lime open squares. Only a few such stars are known. They were reported in the Galactic bulge and disk fields by  \cite{soszynski2019}. One more such star was reported by \cite{jurcsik2015} in the M3 globular cluster.


An interesting group was reported by \cite{smolec.prudil2016} and is plotted in Fig.~\ref{fig:pet_rrl_all} with magenta diamonds. In these stars, the dominant mode of pulsation is the fundamental mode (RRab). The additional signal has a lower amplitude and shorter period, which puts these stars in the long-period extension of the classical RRd sequence. Based on the light curve shapes and theoretical models \cite{smolec.prudil2016} concluded that the double-mode fundamental and first-overtone pulsations are the most probable explanation of periodicities observed in these stars.

\subsection{\rr}
The most common group of multi-periodic stars where the additional signal cannot correspond to another radial mode is plotted with green points in Fig.~\ref{fig:pet_rrl_all}. These are so-called \rr~stars. The dominant mode of pulsation is the first-overtone, so these are either first-overtone (RRc) stars or RRd stars. The additional signal, $f_{0.61}$, has a low amplitude, a period shorter than the first-overtone period, and forms a period ratio of around 0.61 with the first overtone. This period ratio instantly excludes the possibility that the origin of the signal is another radial mode \citep[see e.g. fig. 4 in][]{netzel1}. 

The first identified \rr~star was an AQ Leo \citep{gruberbauer2007}. Currently, more than one thousand such stars are known. They were detected in various systems, both from space and ground-based observations \citep[see e.g.][and references therein]{moskalik2015,jurcsik2015,smolec2017,netzel_census,molnar2022}. At a closer look, the \rr~stars form three sequences in the Petersen diagram at the period ratio of around 0.61, 0.62, and 0.63. 

A common feature of the additional signals in \rr~stars is the detection of signals at $0.5f_{0.61}$ and $1.5f_{0.61}$. We note, that the additional signals, as well as signals $0.5f_{0.61}$ and $1.5f_{0.61}$, are typically non-stationary. This results in wide and complex structures in frequency spectra often offset from the exact half-integer frequency \citep[see e.g.][]{netzel_census}. An interesting group in the Petersen diagram are stars marked with black crosses in Fig.~\ref{fig:pet_rrl_all} located at the period ratio of around 0.80. In these stars, in contrast to stars plotted with lime open squares in Fig.~\ref{fig:pet_rrl_all}, the additional signal has a longer period than the first overtone. A possible scenario explains these signals as subharmonics at $0.5f_{0.61}$ in RR$_{0.61}$ stars \citep{netzel_census,benko2021}.

The explanation of the origin of additional signals in \rr~stars was proposed by \cite{dziembowski2016}. He suggested that the signal that forms period ratio of around 0.61 is caused by harmonics of non-radial modes of degrees $\ell=8$ or 9 for the top and the bottom sequence. The middle sequence corresponds to linear combination of the two other signals.

This form of pulsation seems to be common for first-overtone RR Lyrae stars. The incidence rate of \rr~stars increases with the increasing quality of photometry. In the ground-based studies, the incidence rate of detection typically is below 60\% (the only exception is the study of the globular cluster NGC~6362, where the derived incidence rate is 63\% \citet{smolec2017}). The incidence rate based on the space-based photometry typically exceeds 60\% \citep{molnar2015,moskalik2015,molnar2022,forro2022}. The record holder is four RRc stars observed in the original {\it Kepler} field and studied by \cite{moskalik2015}. In all of them, the additional signal was detected giving the incidence rate of 100\%. Interestingly, a recent study of the background stars in the {\it Kepler} field, which resulted in the detection of four new RRc stars and two RRd stars did not change the incidence rate of \rr~stars \citep{forro2022}. On the other hand, based on the color-magnitude diagrams, the conclusion is that \rr~stars are not detected in its bluest part \citep{jurcsik2015,smolec2017,molnar2022}. Recently \cite{netzel2022} calculated a grid of theoretical models for pulsations in non-radial modes of degrees $\ell=8,9$ in RR Lyrae stars and showed that indeed those modes are linearly stable close to the blue edge of the instability strip.

\subsection{\rrl}

Another puzzling group, where the additional signal cannot correspond to another radial mode are \rrl~stars (orange crosses in Fig.~\ref{fig:pet_rrl_all}). Again, the dominant mode of pulsation is the first overtone. The additional low-amplitude signal has a period longer than the first-overtone period and also longer than the (undetected) fundamental mode. It forms a period ratio of around 0.68 with the first overtone. \rrl~stars were detected both from the ground-based photometry \citep{netzel068,netzel_census} and from the space-based data \citep[][]{moskalik2015,molnar2022}.

\begin{figure*}
    \centering
    \includegraphics[width=0.9\textwidth]{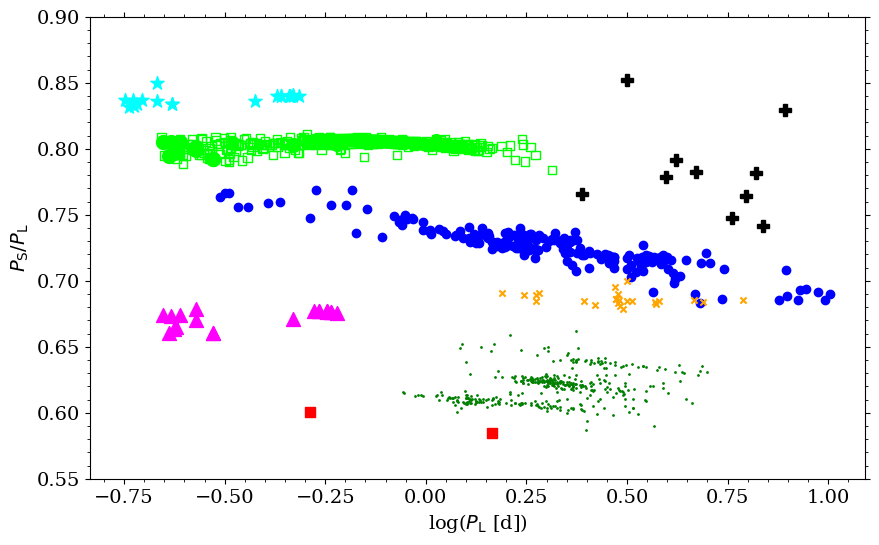}
    \caption{Petersen diagram for multi-mode classical Cepheids.}
    \label{fig:pet_cep_all}
\end{figure*}

The only explanation of the nature of signals in \rrl~stars was proposed by \cite{dziembowski2016}. The observed period ratio of around 0.68 was reproduced assuming that the \rrl~stars are not RR Lyrae stars but stripped giants of masses around 0.2--0.25 \ms~ similar to Binary Evolution Pulsator \citep[BEP,][]{pietrzynski2012}. In this case, the period ratio would correspond to double-mode pulsations in fundamental mode and first overtone. However, there are stars that show both signals that correspond simultaneously to \rrl~and \rr~stars \citep{moskalik2015,netzel_census}. Hence, a consistend and coherent explanation for both groups has still to be found.

\subsection{Peculiar RRd stars (pRRd)}

Another puzzling group of RR Lyrae stars is plotted with cyan triangles in Fig.~\ref{fig:pet_rrl_all}. The dominant mode in the majority of these stars is the fundamental mode, but in a few stars, the dominant mode is the first overtone. The additional signal has a shorter period and forms a period ratio with the dominant frequency around 0.70, which puts this group of stars in the short-period extension on the classical RRd sequence. It was first reported by \cite{prudil2017}. More such stars were detected by \cite{soszynski2019} and \cite{nemec.moskalik2021}. The Fourier coefficients, describing the light curve shape, for these stars are in between values typical for RRc and RRab stars. Using theoretical models \cite{prudil2017} showed that this group of stars cannot be reproduced with double-mode pulsations in fundamental mode and first overtone neither in RR Lyrae stars nor low-mass stripped red giants similar to BEPs. \cite{nemec.moskalik2021} named these stars peculiar RRd stars (pRRd) and showed that their properties are indeed different than for classical or anomalous RRd stars. Presently, we do not know the explanation for the origin of signals in pRRd stars.

\section{Petersen diagram for classical Cepheids}\label{sec:cep}

Petersen diagram for multi-mode classical Cepheids is presented in Fig.~\ref{fig:pet_cep_all}. 

\subsection{Multi-mode radial pulsations}
Double-mode pulsations in classical Cepheids are known for more than fifty years \citep{oosterhoff1957a,oosterhoff1957b} and are relatively common. Currently, we know stars that show double-mode pulsations in fundamental mode and first overtone, first and second overtones, and first and third overtones. These are plotted in Fig.~\ref{fig:pet_cep_all} with blue circles, lime squares, and magenta triangles, respectively. \cite{soszynski2015_blazhko} reported a double-mode Cepheid pulsating in second and third overtones (cyan stars in Fig.~\ref{fig:pet_cep_all}). Triple-mode Cepheids pulsating in radial modes are also known. They either pulsate in fundamental mode, first and second overtones, or in first, second, and third overtones \citep{soszynski2015_blazhko}.

\subsection{0.61 stars}
The analogous group to \rr~stars is known among classical Cepheids \citep{moskalik2008,moskalik2009,soszynski2008}. They are plotted with green points in Fig.~\ref{fig:pet_cep_all}. Again, stars in this group have the first overtone as a dominant mode of pulsation, and the shorter-period signal forms the characteristic period ratio of around 0.61 with the first overtone. Currently, a few hundred such stars are known \citep{soszynski2010,soszynski2015,suveges.anderson2018,smolec.sniegowska2016,rajeev} in both Magellanic Clouds and Galactic disk. These stars also form characteristic three sequences in the Petersen diagram. According to the explanation by \cite{dziembowski2016}, the observed sequences can be explained with harmonics of non-radial modes of degrees $\ell=7$, 8, and 9 for the top, middle, and the bottom sequence, respectively.

A detailed study by \cite{smolec.sniegowska2016} resulted in a discovery of power excess located at a half-integer frequency (around $0.5f_{0.61}$). Interestingly, stars showing this power excess are typically members of the middle sequence in the Petersen diagram. This is consistent with the predictions of model by \cite{dziembowski2016}. Namely, non-radial modes of $\ell=8$ are least affected by the cancellation effect. For details see a discussion in \cite{smolec.sniegowska2016} and a similar discussion for RR Lyrae stars in \cite{netzel_census}.

A group of stars plotted in Fig.~\ref{fig:pet_cep_all} with black pluses was reported only recently by \cite{rajeev}. This group is analogous to RR Lyrae stars plotted in Fig.~\ref{fig:pet_rrl_all} with black pluses.

Interestingly, the $f_{0.61}$ signal was recently detected in an anomalous Cepheid XZ Cet during the analysis of the TESS data by \cite{plachy2021}. This is the first and only such discovery in anomalous Cepheids so far.

\subsection{0.68 stars}

\cite{suveges.anderson2018} performed an analysis of additional periodicities in classical Cepheids in Magellanic Clouds. They reported the discovery of stars that are counterparts of RR$_{0.68}$ group known among the RR Lyrae stars. These stars are plotted in Fig.~\ref{fig:pet_cep_all} with orange crosses. The very first member of this group is CoRoT~0223989566, which is a double mode 1O/2O Cepheid with an additional long-period signal that forms a period ratio of 0.682 with the first overtone \citep{poretti2014}. However, with only one star, no new multi-periodic group was established.

\section{Blazhko effect}\label{sec:blazhko}

The Blazhko effect is a quasi-periodic modulation of phase and/or amplitude of pulsation. Since its discovery by \cite{blazhko} more than a century ago its physical origin is still unknown. However, our understanding of the observed properties of Blazhko stars grows significantly. 

If the amplitude of modulation is big enough, the modulation is visible in light curve \citep[see e.g. fig. 4][]{benko2014}. In frequency spectra, it manifests as equidistant sidepeaks centered at the dominant frequency and its harmonics. 


The Blazhko effect is observed in RRab and RRc stars. It is more common among RRab stars. The incidence rate of Blazhko RRab stars among RRab stars is typically around 50\%. Interestingly, this value is reached by studies that incorporate ground-based observations as well as space-based photometry \citep[see e.g.][]{jurcsik2014,prudil.skarka2017,molnar2022}. On the other hand, \cite{kovacs2018} analyzed the {\it Kepler} photometry from the K2 mission and reported an incidence rate of 91\% suggesting the underlying incidence rate of 100\%. In particular, \cite{kovacs2018} reported significantly more stars with low-amplitude modulation. \cite{benko2019} studied non-Blazhko RRab stars in the original {\it Kepler} field and detected subtle cycle-to-cycle variations of light curve (see their fig. 1). Such variations can result in residual signals close to the fundamental frequency. So, the incidence rate of Blazhko RRab stars remains a debated
topic. The incidence rate of RRc stars showing the Blazhko modulation is lower. \cite{netzel_blazhko} reported only 5.6\%  based on the analysis of the ground-based data. Based on the space-based TESS data \cite{molnar2022} reported an incidence rate of 13\%, which is still significantly smaller than the value inferred for RRab stars.

The modulation is also observed in double-mode RR Lyrae stars, namely in aRRd \citep{jurcsik2014,soszynski2014,smolec2015,plachy2017rrdbl,soszynski2016}. These stars have also peculiar properties of pulsations as discussed in Sec.~\ref{sec:rrl}. Interestingly, sometimes both modes are modulated, but sometimes only the fundamental mode or first overtone can be modulated. As suggested by \cite{soszynski2016}, the resonance $2f_{\rm 1O}\approx f_{\rm F}+f_{\rm 2O}$ might be involved in observed pulsation properties of aRRd stars.

The Blazhko modulation was detected in classical Cepheids as well. \cite{smolec2017_cep} studied the most numerous sample of fundamental-mode Cepheids in the Magellanic Clouds and detected modulation in over 50 Cepheids. Even though the overall incidence rate is small, it is high for fundamental periods around 10 to 15\,d. Different incidence rates for Small and Large Magellanic Cloud suggest that the modulation depends on the metallicity. Interestingly, \cite{smolec2017_cep} reported a characteristic modulation period, which seems to be around ten times the fundamental-mode period. \cite{soszynski2015_blazhko} reported a modulation visible in the light curve of a first-overtone classical Cepheid. The modulation was also detected in a second-overtone Cepheid, V473 \citep{molnar.szabados2014,molnar2017_v473}.

\cite{moskalik.kolaczkowski2009} reported a modulation in double-mode classical Cepheids pulsating in first and second overtones. It was detected in 19\% of the studied sample. Interestingly, both modes are modulated with the same long period. Amplitude modulation of both modes is anticorrelated.

Only recently \cite{rajeev} reported modulation in double-mode fundamental and first-overtone Cepheids.

\subsection{Explaining the Blazhko effect}
Any model attempting to explain the Blazhko effect should take into account the observed properties of the Blazhko effect. That includes the incidence rates, which are different for different types of stars as discussed in the previous section. Apart from the incidence rates, Blazhko stars show a variety of characteristics. Modulation periods vary from very short, from below 5\,d \citep{skarka2020,netzel_blazhko}, to very long, of the order of $10^4$\,d \citep{jurcsik.smitola2016,netzel_blazhko,skarka2020}. However, the upper limit, if exists, is uncertain, because it depends on the length of data. 

Complex modulation patterns are relatively often observed. That includes multiple modulation periods or multiplet structures beyond simple dublets and triplets \citep[see e.g.][]{chadid2010,guggenberger2012,netzel_blazhko}. The sidepeaks on the lower and higher frequency side can show asymmetry in amplitudes \citep[see theoretical discussion in][]{benko2011}. In extreme cases, sidepeak on one side can be hidden in the noise. In this scenario only one sidepeak is detected, instead of multiplets centered at pulsation frequency and its harmonics \citep[see a discussion in ][]{netzel_blazhko}. With the excellent data quality of space-based photometry, it is possible to detect the low-amplitude sidepeaks as well. For instance, \cite{molnar2022} from their analysis of TESS photometry, reported Blazhko stars that show extreme amplitude asymmetry. 

Long-term monitoring of Blazhko stars also showed that the Blazhko effect itself can change over time. This irregularity can manifest as cycle-to-cycle variations \citep[e.g.][]{guggenberger2011} or as slow long-term trends \citep{leborgne2014}.

Many models have been proposed to explain the Blazhko effect, but none of them yet fully succeeded to do that \citep[for details see ][and references therein]{smolec2016}. That includes magnetic oblique rotator/pulsator model \citep{shibahashi2000}, non-radial resonant model \citep{nowakowski.dziembowski2001}, magnetic dynamo-driven convection \citep{stothers2006}, atmospheric shocks' dynamics \citep{gillet2013} and non-resonant radial non-radial mode interactions \citep{cox2013}. 

The continuous photometric observations by the {\it Kepler} satellite led to the discovery of the period-doubling phenomenon in Blazhko RRab star \citep{kolenberg2010}. In the light curve, it shows as alternating minima and maxima. In the frequency spectrum, it causes the presence of subharmonics, i.e. signals at half-integer frequencies. Interestingly, the period-doubling was never detected in Blazhko RRc stars. \cite{szabo2010} suggested, that the period doubling is due to 9:2 resonance between the fundamental mode and the 9th radial overtone. This motivated new efforts in modeling. \cite{buchler.kollath2011} proposed that the 9:2 resonance causing the period doubling might also be responsible for the Blazhko effect. Recently, \cite{kollath2021} was able to produce amplitude modulation and period doubling in a model of an RR Lyrae star, though yet not able
to explain all the observed aspects of the Blazhko effect.

\begin{acknowledgements}
I would like to thank the organizers of the RRL conference for organizing a wonderful meeting and giving me the opportunity to present these exciting updates to the community.

H.N.\ has been supported by the Lend\"ulet Program of the Hungarian Academy of Sciences, project No. LP2018-7/2022 and by the \'UNKP-22-4 ``Tudom\'annyal Fel!" New National Excellence Program of the Ministry of Innovation and Technology from the source of the National Research, Development and Innovation Fund. 

I am grateful to R. Szabo for reading the manuscript.

\end{acknowledgements}

\bibliographystyle{aa}
\bibliography{bibliography}

\end{document}